\documentclass{article}
\usepackage{arxiv}

\usepackage[T1]{fontenc}    
\usepackage{amsmath}
\usepackage{amssymb}
\usepackage{booktabs}       
\usepackage{amsfonts}       
\usepackage{nicefrac}       
\usepackage{microtype}      
\usepackage{lipsum}		
\usepackage{graphicx}
\usepackage{doi}
\usepackage{subfiles}
\usepackage[linesnumbered,lined,boxed,commentsnumbered]{algorithm2e}
\usepackage{caption}
\usepackage{subcaption}
\usepackage{cancel}

\title{Modeling Randomly Walking Volatility \\ with Chained Gamma Distributions}

\author{ 
	{Di Zhang\textsuperscript{a}, Qiang Niu\textsuperscript{b}, Youzhou Zhou\textsuperscript{b}} \\
	{School of AI and Advanced Computing\textsuperscript{a}, School of Mathematics and Physics\textsuperscript{b}}\\
	Xi'an Jiaotong-Liverpool University\\
	Suzhou, 215123, China PR \\
	\texttt{\{di.zhang, qiang.niu, youzhou.zhou\}@xjtlu.edu.cn}
}

\date{}

\renewcommand{\headeright}{}
\renewcommand{\undertitle}{}
\renewcommand{\shorttitle}{Modeling Randomly Walking Volatility with Chained Gamma Distributions}

\begin{document}
	\maketitle
	\begin{abstract}
Volatility clustering is a common phenomenon in financial time series. Typically, linear models can be used to describe the temporal autocorrelation of the (logarithmic) variance of returns. Considering the difficulty in estimating this model, we construct a Dynamic Bayesian Network, which utilizes the conjugate prior relation of normal-gamma and gamma-gamma, so that its posterior form locally remains unchanged at each node. This makes it possible to find approximate solutions using variational methods quickly. Furthermore, we ensure that the volatility expressed by the model is an independent incremental process after inserting dummy gamma nodes between adjacent time steps. We have found that this model has two advantages: 1) It can be proved that it can express heavier tails than Gaussians, i.e., have positive excess kurtosis, compared to popular linear models. 2) If the variational inference(VI) is used for state estimation, it runs much faster than Monte Carlo(MC) methods since the calculation of the posterior uses only basic arithmetic operations. And its convergence process is deterministic.

We tested the model, named Gam-Chain, using recent Crypto, Nasdaq, and Forex records of varying resolutions. The results show that: 1) In the same case of using MC, this model can achieve comparable state estimation results with the regular lognormal chain. 2) In the case of only using VI, this model can obtain accuracy that are slightly worse than MC, but still acceptable in practice; 3) Only using VI, the running time of Gam-Chain, in general case, can be reduced to below 5\% of that based on the lognormal chain via MC.
	\end{abstract}
	\keywords{Stochastic Volatility \and Variational Inference \and Dynamic Bayesian Network \and Cryptocurrency}
	
	\section{Introduction}
	
	In financial markets, asset prices are constantly fluctuating. The strength of volatility is usually expressed in terms of the logarithmic variance\footnote{Variance, standard deviation, or logarithmic standard deviation are also used in other literature. For brevity, this paper will use 'volatility' to denote the 'logarithmic variance' uniformly.} of returns\footnote{i.e., rate of returns.}. Most financial applications, such as risk management, derivatives pricing, portfolio management, etc., require a reasonable estimation of the current volatility. After long-term observations, it has been found that: 1) The volatility embedded in the series does not remain constant but seems to follow another distribution, which leads to the phenomenon of heavy tails of returns; 2) the distribution behind volatility is not fixed, but changes over time and exhibits a certain degree of positive autocorrelation.
	
	A popular model for this phenomenon is Stochastic Volatility\cite{Andersen2009StochasticV} (SV). It first defines an unobservable stochastic process that expresses the change in volatility, then makes it instant variances of returns, and finally generates a sequence of observable values. Suppose the change in volatility is defined via a linear Gaussian model, such as Autoregressive Moving Average Model (ARMA). In that case, this approach yields understandable results and always reliably converges in computation. However, it also has two shortcomings: 1) through the observation of actual data, we find that the increment of volatility does not necessarily obey a Gaussian distribution, and it is also heavy-tailed in most cases; 2) when volatility is estimated, the linear Gaussian model has no closed-form posterior. Therefore, it needs to depend, or partly depend on sampling, to express the volatility distribution in the form of a bunch of particles. This sampling process often consumes much time. Moreover, when it works with parameter estimation in Expectation Maximization (EM), it is not easy to judge whether the whole process has converged.
	
	To this end, we reconsider this problem from the Dynamic Bayesian Network (DBN) perspective. We note that the usual SV model can be viewed as a two-layer state space model \cite{Bishop_2006}, where the change in volatility is defined in the transition equation, and the price change is defined in the observation equation. To make this model easier to compute, we define it as follows: 1) Note that for the rate parameter of a gamma distribution, its conjugate prior is also a gamma distribution. Therefore, we first use a series of inter-connected gamma distributions to represent volatility. 2) Through such a direct connection, the obtained volatility increment is not independent but depends on the absolute value of the last step. However, the increments between every two steps are independent, similar to a random walk. 3) At every interval, let the volatility correspond to the first gamma distribution by a prior of the precision (=1/variance) of a normal distribution. (Or equivalently, we are inserting 'dummy node' between adjacent steps.) 4) The set of all random variables corresponding to normal distributions constitutes the return sequence, which can be observed.
	
	The primary advantage of this model is computation. At each node of this model, the posterior maintains the form of the gamma distribution, which makes the approximate estimation of states relatively fast. It may take multiple scans of the sequence to converge, instead of only two passes are needed by the forward-backward approach commonly used in DBNs. However, this is not a problem in practice. Because the model parameters cannot be manually set in advance, the state estimation is often performed inside the iteration of the EM algorithm; even if the state estimation only scans the sequence once per round, it will converge to the (local) optimum with model parameters eventually. Moreover, the items that do not change in each scan can be calculated outside the EM's loop in advance, further speeding up the process (Sec. \ref{sec:algo}).
	
	The significance of this model is more than the calculation. It can be shown that this construction implements a random walk of volatility. Like the usual Gaussian random walk, its incremental offset is $0$, and the variance can also take values between $(0,\infty)$. However, the kurtosis it can reach is between $(3,6)$, which is larger than the $3$ of the Gaussian random walk, which means that, to some extent, it can express the heavy tail of the volatility increment. The heavy-tail inside volatility exists, and it has been discussed in empirical research \cite{carnero2004persistence}. This supports the view that our model provides at least a different option, which does not need additional random variables to express but inherently incorporates this effect.
	
	We have tested the Gam-Chain model in different markets at different resolutions. Experiments show that although the VI method can only get an approximate solution for volatility, it can scale a significant part of instruments' returns to the standard normal distribution, benefiting from the heavy tail of the gamma distribution. As a comparison, if we generate the posterior based on the lognormal chain, it will be difficult to use directly, for its tail is too thin, and thus the normalization effect is poor. Although there is no significant difference in the algorithmic complexity of the VI method compared to MC, it can run faster since it does not require sampling during state estimation but only basic arithmetic operations ($+$, $-$, $\times$ and $\div$). In addition, it is easy to judge convergence because the calculation process is entirely deterministic.
	
\section{Related Work}

\subsection{Volatility Clustering}

The phenomenon of volatility clustering was first discussed in \cite{mandelbrot1963variation}, which mentioned: "large changes tend to be followed by large changes, of either sign, and small changes tend to be followed by small changes." \cite{Cont2001EmpiricalPO} lists volatility clustering as one of the "stylized facts" and believes that it is common in securities markets in different periods and countries. \cite{lux2000volatility} uses an artificial market to simulate this phenomenon by placing a certain percentage of chartists and fundamentalists. In this market, when the price crosses a certain threshold, the chartist's trading causes volatility to explode, but the fundamentalist gradually redirects it towards stability.

Two popular classes of models have been used to describe this phenomenon: the Generalized Autoregressive Conditional Heteroskedasticity (GARCH) model\cite{bollerslev1986generalized} and the SV model\cite{heston1993closed}. The common thing is that they describe the current volatility as a function of past volatilities. If the function is ARMA and is entirely deterministic (i.e., the noise is zero), then it is GARCH; if the function itself is derived from another random process, then it is SV. The latter includes a large class of models, and some popular variants are introduced in \cite{Andersen2009StochasticV}, including discrete and continuous, linear and nonlinear, etc. In estimation methods, GARCH usually uses a two-step MLE method, first estimating the residual of returns and then estimating the ARMA coefficients for the volatility, while SV generally needs to be estimated using pseudo-likelihood or Markov chain Monte Carlo (MCMC)\cite{Broto2004EstimationMF}.

The models and their variants discussed in this article (Sec. \ref{sec:gam} and \ref{sec:var}) is probably one of the simplest forms of expressing SV by directly assuming that volatility follows a random walk (Eq. \ref{eqn:frame}). The reasons for restricting this form are: 1) this paper can focus the discussion on the core idea and leave extensions to possible future works; 2) it is probably enough for many applications where overfitting caused by excessive parameters is undesirable\cite{pymc3}.

\subsection{Dynamic Bayesian Network}

SV can be naturally represented as DBN. Work in this area can be traced back to \cite{Jacquier1994BayesianAO}, which uses MCMC for model estimation. Among them, we pay special attention to applying variational methods to this problem. There are mainly two approaches here: 1) Keep the form of SV unchanged, in which the variational method is used to obtain an approximate solution, and then the sampling is performed. For example, \cite{Kleppe2012FittingGS} uses Laplace approximation (LA) to generate proposal distributions to improve sampling efficiency. Further, it is possible to cancel the MC process and directly use nested LA for approximation \cite{Bermudez2021IntegratedNL}. 2) Change the form of SV to make it more suitable for VI. A key element here is the gamma distribution. The variance-gamma distribution can be obtained by directly using the gamma distribution to represent the variance and compounding it with the normal distribution \cite{Madan1990TheVG}. Subsequently, \cite{Gelman2004PriorDF} suggested switching to an inverse gamma distribution (or equivalently, with all returns' precision (=1/sigma) distributed as a gamma) in order to keep the posterior's form unchanged in Bayesian inference. Further, \cite{Langren2015SwitchingTN} uses the inverse gamma process to describe the variance of the variance for better option pricing results. Based on the conjugate relationship of inverse gamma and normal, \cite{LenGonzlez2018EfficientBI} samples directly from the posterior to obtain estimates of fluctuations. \cite{Santos2018ABG} and \cite{Rezende2022ANF} change the observation distribution from normal to Generalized Error Distribution (GED), which still uses the gamma distribution for its precision so that the likelihood can be marginalized and operated in closed form. The heavy tail inside variance can also be expressed.

The difference in this paper is that we directly use the nonlinear gamma chain to express the fluctuation. We no longer resort to other more complex or indirect methods.
	
	\section{Gamma Chain Model}
	
\subsection{Problem Statement}

Consider a SV model of the form:
\begin{equation}\label{eqn:frame}
	\begin{aligned}
		\boldsymbol{y}_t-\boldsymbol{y}_{t-1} &\sim N(0,\boldsymbol{u}_t^{-1}), \\
		\log \boldsymbol{u}_t-\log \boldsymbol{u}_{t-1}  &\sim f^*(\boldsymbol{\theta})
	\end{aligned}
\end{equation}
Among them, $\{\boldsymbol{y}_t\}$ is the observation sequence, $N$ is the normal distribution (for the convenience of the variational derivation in Sec. \ref{sec:se}, the variance here is set as the reciprocal of $\boldsymbol{u}_t$), $\{\boldsymbol{u}_t\}$ is the volatility\footnote{Different from above, it is actually logarithmic precision, i.e. negative logarithmic variance . This setting is only for convenience and does not affect our final result. } sequence (unobservable), $f^*$ is a probability distribution function. It depends only on its parameter $\boldsymbol{\theta}$ and does not change over time. Obviously, $\{\log\boldsymbol{u}_t\}$ has the property of independent increment, and $\{\Delta\log\boldsymbol{u}_t\}$ is stationary. when
$f^*(\boldsymbol{\theta})\triangleq N(0,S^2)$
, Eq. \ref{eqn:frame} is
\begin{equation}\label{eqn:logn}
	\begin{aligned}
		\boldsymbol{y}_t-\boldsymbol{y}_{t-1} &\sim N(0,\boldsymbol{u}_t^{-1}), \\
		\boldsymbol{u}_t &\sim LogN(\log \boldsymbol{u}_{t-1},S^2)
	\end{aligned}
\end{equation}
where LogN is lognormal distribution. Here $\{\log \boldsymbol{u}_t\}$ obeys a Gaussian random walk process. Note that the variance of $\{\Delta\log \boldsymbol{u}_t\}$ is $S^2$ and the kurtosis is $3$.

The questions to be studied in this section is, 1) Can a new $f^*$ be defined such that the kurtosis of $\{\Delta\log\boldsymbol{u}_t\}$ is greater than 3? 2) How to quickly estimate ${\boldsymbol{u}_t}$ for a given observation ${\boldsymbol{y}_{1:T}}$; 3) Estimation of parameter $\boldsymbol{\theta}$.
	
\subsection{Model Definition}\label{sec:gam}

As a preliminary attempt, we tentatively use straightforwardly a gamma chain to express the change in volatility, which is
\begin{equation}\label{eqn:naive}
	\boldsymbol{u}_t \sim   Ga(A,\boldsymbol{u}_{t-1})
\end{equation}
Where $Ga(A,\boldsymbol{u}_{t-1})$ represents the gamma distribution with shape $A$ and rate $\boldsymbol{u}_{t-1}$. To align Eq. \ref{eqn:naive} with Eq. \ref{eqn:frame}, we denote $\boldsymbol{w}_t\triangleq \Delta\log(\boldsymbol{u}_t)$, then Eq. \ref{eqn:naive} is rewritten as (details in \ref{sec:dev_inc})
\begin{equation}\label{eqn:naive2}
	p(\boldsymbol{w}_t) =\frac{e^{ -e^{\boldsymbol{w}_t}\boldsymbol{u}_{t-1}^2} \left(e^{\boldsymbol{w}_t}\boldsymbol{u}_{t-1}^2 \right)^A}{\Gamma (A)}
\end{equation}
where $\Gamma$ is the gamma function. Note that the distribution parameter in Eq. \ref{eqn:naive2} is $(A,\boldsymbol{u}_{t-1})$, where $\boldsymbol{u}_{t-1}$ is obviously not time-invariant, and does not meet the requirements of $f^*$ in Eq. \ref{eqn:frame}.

\begin{figure}[htbp]
	\centering
	\includegraphics[width=0.55\linewidth]{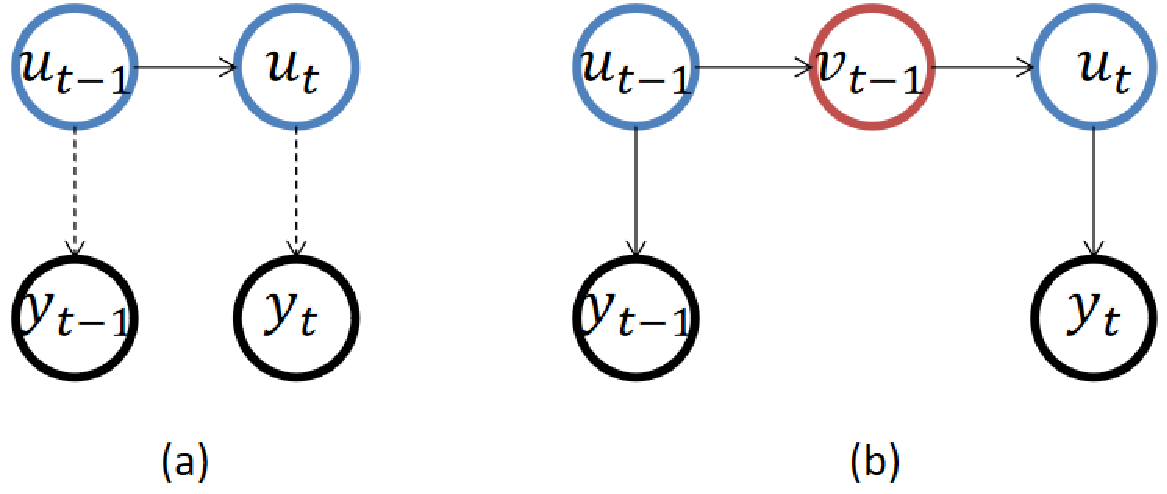}
	\caption{Compare the lognormal distribution chain (a) and the gamma distribution chain (b). Black represents observations, blue represents fluctuations, and red represents dummy nodes. The solid line represents a closed-form local posterior, and the dashed line represents that the closed-form no longer exists after adding that node.}
	\label{fig:design}
\end{figure}
To fix this defect, we insert a gamma-distributed random variable $\boldsymbol{v}_t$ after each $\boldsymbol{u}_t$ (refer to Fig. \ref{fig:design}) to construct a random process as
\begin{equation}\label{eqn:fix}
	\boxed{
		\begin{aligned}
			\boldsymbol{y}_t-\boldsymbol{y}_{t-1} &\sim N(0,\boldsymbol{u}_t^{-1}),\\
			\boldsymbol{u}_t &\sim   Ga(A,\boldsymbol{v}_{t-1}), \\
			\boldsymbol{v}_{t-1} &\sim Ga(A,\boldsymbol{u}_{t-1})
		\end{aligned}
	}
\end{equation}
We marginalize out $\boldsymbol{v}_{t-1}$ and align it again to the Eq. \ref{eqn:frame} (details in \ref{sec:dev_inc}):
\begin{equation}\label{eqn:fix2}
	\begin{aligned}
		p(\boldsymbol{w}_t) = \frac{e^{-A \boldsymbol{w}_t} \left(e^{\boldsymbol{w}_t}+1\right)^{-2 A} \Gamma (2 A)}{\Gamma (A)^2}
	\end{aligned}.
\end{equation}
$\boldsymbol{u}_{t-1}$ no longer exists here because it has been eliminated, and only parameter $A$ remains. Thus we can define a qualified $f^*$ as the right side of Eq. \ref{eqn:fix2}. Next, the variance and kurtosis are calculated to compare the expressivity with Eq. \ref{eqn:logn}. The moment generation function of Eq. \ref{eqn:fix2} is (details in \ref{sec:dev_mgf}):
\begin{equation}\label{eqn:mgf}
	\phi(\lambda)=\mathbb{E}[e^{\lambda \boldsymbol{w}_t}]=\frac{\Gamma (A-\lambda) \Gamma (A+\lambda)}{\Gamma (A)^2}.
\end{equation}
Deriving Eq. \ref{eqn:mgf} and calculating the central moments from the 1st to 4th order, its variance can be obtained as
\begin{equation}\label{eqn:var}
	V=Var(\boldsymbol{w}_t)=2 \psi^{(1)}(A)
\end{equation}
, and kurtosis as (details in \ref{sec:dev_kurt})
\begin{equation}\label{eqn:kurt}
	K=Kurt(\boldsymbol{w}_t)=3 + \frac{\psi^{(3)}(A)}{2 \left[\psi^{(1)}(A)\right]^2}
\end{equation}
Where $\psi$ is the digamma function. Note that $\underset{A\to 0}{\text{lim}}V=\infty$, $\underset{A\to \infty }{\text{lim}}V=0$, thus the variance can be assume to $(0,\infty)$, which is equivalent to the expressive power of Eq. \ref{eqn:logn}. The value range of kurtosis is wider, such as $\underset{A\to 0}{\text{lim}}K=6$, $\underset{A\to \infty }{\text{lim}} K=3$. It can be shown that the kurtosis of Eq. \ref{eqn:fix2} is exactly within the interval (3,6) (details in \ref{sec:proof_kurt}).

\subsection{State Estimation}\label{sec:se}

Under the Bayesian perspective, $\boldsymbol{z}\triangleq\{\boldsymbol{u},\boldsymbol{v}\}_{1:T}$ is the set of all state variables. State estimation is to find the posterior $p(\boldsymbol{z}\mid\boldsymbol{y})$ under the given observation $\boldsymbol{y}\triangleq\{\boldsymbol{y}_{1:T}\}$. It should be pointed out that neither the model of Eq. \ref{eqn:logn} nor the model of Eq. \ref{eqn:fix} can analytically give an exact posterior. However, for the Eq. \ref{eqn:fix}, at each local $p(\boldsymbol{u}_t\mid\boldsymbol{v}_{t},\boldsymbol{y}_t,\boldsymbol{v} _{t+1})$, $p(\boldsymbol{v}_t\mid\boldsymbol{u}_{t-1},\boldsymbol{u}_t)$, yet it can give the exact posterior, and keep the form of the gamma distribution unchanged. Thus, we can use variational inference to find every local analytical solution and iteratively find a global approximate solution.

Based on variational inference, we set the optimization objective to maximize the loss function
\begin{equation}\label{eqn:target}
	\mathcal{L}(q)=\int q(\boldsymbol{z})\ln \{ \frac{p(\boldsymbol{z},\boldsymbol{y})}{q(\boldsymbol{z})} \} \text{d} \boldsymbol{z}.
\end{equation}
where $q$ is the posterior probability to be solved. Since finding an exact solution for the formula \ref{eqn:target} is still difficult, we further give the mean filed assumption\cite{Bishop_2006}:
\begin{equation}\label{eqn:q}
	q(\boldsymbol{z})=\prod_{i=1}^{2 T} q_i(\boldsymbol{z}_i)
\end{equation}
After substituting Eq. \ref{eqn:q} into Eq. \ref{eqn:target}, then for each $q_i$, the optimal solution should satisfy\cite{Bishop_2006}
\begin{equation}\label{eqn:q2}
	q_i^*(\boldsymbol{z}_i)=\frac{\exp \{ \mathbb{E}_{j\neq i}[\ln p(\boldsymbol{z},\boldsymbol{y})] \} }
	{\int \exp \{ \mathbb{E}_{j\neq i}[\ln p(\boldsymbol{z},\boldsymbol{y})] \} \text{d}\boldsymbol{z}_i }
\end{equation}.
Then, substitute Eq. \ref{eqn:frame} and Eq. \ref{eqn:fix} into Eq. \ref{eqn:q2}. and simplify it. We can see that $q_i^*(\boldsymbol{z}_i) \sim Ga(\boldsymbol{a}_i,\boldsymbol{b}_i)$, and their parameters are
\begin{equation}\label{eqn:u}
	\boldsymbol{a}_t^{(u)},\boldsymbol{b}_t^{(u)}=\left\{\begin{matrix}
		A+3/2,& \boldsymbol{a}_{t+1}^{(v)}/\boldsymbol{b}_{t+1}^{(v)}+(\Delta\boldsymbol{y}_{t})^2/2& & for &t=1\\
		\\
		2 A+1/2,& \boldsymbol{a}_{t}^{(v)}/\boldsymbol{b}_{t}^{(v)}+\boldsymbol{a}_{t+1}^{(v)}/\boldsymbol{b}_{t+1}^{(v)}+(\Delta\boldsymbol{y}_{t})^2/2& & for &t>1
	\end{matrix}\right.
\end{equation}
\begin{equation}\label{eqn:v}
	\boldsymbol{a}_t^{(v)},\boldsymbol{b}_t^{(v)}=\left\{\begin{matrix}
		2 A,& \boldsymbol{a}_{t-1}^{(u)}/\boldsymbol{b}_{t-1}^{(u)}+\boldsymbol{a}_{t}^{(u)}/\boldsymbol{b}_{t}^{(u)}& & for &t<T\\
		\\
		A,& \boldsymbol{a}_{t-1}^{(u)}/\boldsymbol{b}_{t-1}^{(u)} && for &t=T
	\end{matrix}\right.
\end{equation}
At the beginning of the iteration, we set the initial value of all $\{\boldsymbol{a}_t,\boldsymbol{b}_t\}$ to $\{1/2,\boldsymbol{y}_{t}^2/2\}$ (that is, the corresponding posterior of one single observation, where the prior is $Ga(0,0)$). After iteratively updating Eq. \ref{eqn:u} and Eq. \ref{eqn:v}, the desired result is obtained after convergence.

\subsection{Parameter Estimation}

Below, we estimate the parameter $A$ by the EM algorithm \cite{Bishop_2006}. In each iteration, it maximizes the following objective (M-step) based on the state estimate (E-step) in Sec. \ref{sec:se}
\begin{equation}\label{eqn:qfunc}
	\begin{aligned}
		\mathcal{Q}(A,A^{\text{(old)}})\triangleq
		\mathbb{E}\left[\mathit{l}(A)\mid\boldsymbol{y},A^{\text{(old)}}\right]
	\end{aligned}
\end{equation}
where $\mathit{l}(A)\triangleq \sum_{t=1}^T\log p(\boldsymbol{u}_t,\boldsymbol{v}_t, \Delta\boldsymbol{y}_t\mid A )$.
Substitute Eq. \ref{eqn:fix} into Eq. \ref{eqn:qfunc}, and get
\begin{equation}\label{eqn:qq}
	\begin{aligned}
		\begin{aligned}
			\mathcal{Q}(A,A^{\text{(old)}})=&
			A \left\{\sum_{t=1}^{T} 2  \mathbb{E}\left[\log \boldsymbol{u}_t\right]+ \mathbb{E}\left[\log\boldsymbol{v}_{t}\right] +\sum_{t=2}^{T} \mathbb{E}\left[ \log \boldsymbol{v}_{t-1}\right]\right\}\\
			&-\sum_{t=1}^{T} 2 \log \Gamma (A)\\
			&+\text{const}.
		\end{aligned}
	\end{aligned}
\end{equation}
Its gradient is
\begin{equation}\label{eqn:grad}
	\begin{aligned}
		\nabla\mathcal{Q}(A,A^{\text{(old)}})=&
		\sum_{t=1}^{T} 2  \mathbb{E}\left[\log \boldsymbol{u}_t\right]+ \mathbb{E}\left[\log\boldsymbol{v}_{t}\right] +\sum_{t=2}^{T} \mathbb{E}\left[ \log \boldsymbol{v}_{t-1}\right]\\
		&-\sum_{t=1}^{T} 2 \psi^{(0)}(A)
	\end{aligned}
\end{equation}
In addition, for the posterior $z\sim Ga(a,b)$ of any hidden state, we have
\begin{equation}\label{eqn:expec}
	\begin{aligned}
		\mathbb{E}[z]&=a/b\\
		\mathbb{E}[\log z]&=\psi^{(0)}(a)-\log b
	\end{aligned}
\end{equation}
Just substitute Eq. \ref{eqn:u} and Eq. \ref{eqn:v} into Eq. \ref{eqn:expec} to calculate the expected expectation of Eq. \ref{eqn:grad} to get the current gradient. At the beginning of each M-step, set the initial value of $A$ to $1$, and then use the gradient ascent method to find $A$ that maximizes Eq. \ref{eqn:qfunc}.

\subsection{Algorithm}\label{sec:algo}

\begin{algorithm}[!htbp]
	
	\SetAlgoLined
	\While{$A$ has not coveraged}{\label{step:loop1}
		\tcp{E Step}
		\For{t=1:T}{
			update $\boldsymbol{b}_t^{(u)}$,$\boldsymbol{b}_t^{(v)}$\label{step:s}
		}
		\For{t=1:T}{
			calculate $\mathbb{E}[\boldsymbol{u}_t],\mathbb{E}[\log\boldsymbol{u}_t],\mathbb{E}[\log\boldsymbol{v}_t]$\label{step:e}
		}
		\BlankLine
		\tcp{M Step}
		\For{t=1:T}{\label{step:mloop}
			calc $\nabla\mathcal{Q}_t(A,A^{\text{(old)}})$\label{step:g}
		}\label{step:mloop_end}
		$A\gets A^{(old)}+\lambda\sum_t\nabla\mathcal{Q}_t(A,A^{\text{(old)}})$ \tcp{gradient ascent}\label{step:a}
	}\label{step:loop1_end}
	\caption{The brief procedure of estimation of the states and parameter for Gam-Chain.}
	\label{alg:pro}
\end{algorithm}
See Algo. \ref{alg:pro} for the program's main process. Note that the running time of
Line \ref{step:s} is $s_{\text{a}}$\footnote{'a' represents arithmetic calculation.}, the running time of line \ref{step:e} is $e$, Line \ref{step:g}-Line \ref{step:a} is $g$. And, the number of iterations of the loop \ref{step:loop1}-\ref{step:loop1_end} is $L$, then the running time of the algorithm is roughly $\mathcal{O}\left(L*T*(s_{\text{a}}+g +e+a)\right)$. During implementation, we should try to put the repeated calculations outside the loop as much as possible. For example, the $(\Delta\boldsymbol{y}_{t})^2/2$ operation in Eq. \ref{eqn:u} actually only needs to be calculated once. For another example, although both Line \ref{step:e} and \ref{step:a} contain $\psi^{(0)}(A)$, it can be extracted up to the outer loop \ref{step:loop1}-\ref{step:loop1_end}, and it complexity will not increase with sequence length.

For each step: Since $g+a$ is usually arithmetic operations, there should be not too much impact on performance; $e$ needs to calculate the logarithm, which is the same whether in this algorithm or Algo. \ref{alg:lcm}. Thus, the key to boosting is $s_{\text{a}}$, and its runtime should be critical. In Sec. \ref{sec:perf} we will give detailed comparisons.

\subsection{Several Variants}\label{sec:var}

\subsubsection{Gam-Chain/MC}\label{sec:var1}
For the algorithm in Sec. \ref{sec:algo}, we name it Gam-Chain/VI. We can also use MC to estimate the model of Eq. \ref{eqn:fix}. Compared with Gam-Chain/VI, it differs only in E-step, where particle smoothing is used for state estimation\cite{godsill2004monte}.
\begin{algorithm}[!htbp]
	
	\SetAlgoLined
	\setcounter{AlgoLine}{2}
	
	\tcp{E Step}
	\tcp{forward sampling}
	
	\For{t=1:T}{
		\For{i=1:N}{\tcp{number of particles}
			$\begin{aligned}
				\boldsymbol{w}_{\boldsymbol{u}_t}^{(i)} &\propto \boldsymbol{w}_{\boldsymbol{v}_{t-1}}^{(i)} p(\boldsymbol{y}_t\mid\boldsymbol{u}_{t}) \\
				\boldsymbol{w}_{\boldsymbol{v}_t}^{(i)}&=\boldsymbol{w}_{\boldsymbol{u}_t}^{(i)}
			\end{aligned}$\tcp{update weights}
		}
	}
	\tcp{backward smoothing}
	\For{t=1:T}{
		\For{i=1:N}{
			$\begin{aligned}
				&\boldsymbol{w}_{\boldsymbol{u}_t\mid\boldsymbol{v}_t}^{(i)} \propto \boldsymbol{w}_{\boldsymbol{u}_{t}}^{(i)} p(\widetilde{\boldsymbol{v}}_t\mid\boldsymbol{u}_{t}^{(i)}) \\
				&\text{choose }\widetilde{\boldsymbol{v}}_{t-1}\text{ with probability }\boldsymbol{w}_{\boldsymbol{u}_t\mid\boldsymbol{v}_t}^{(i)}\\
				&\boldsymbol{w}_{\boldsymbol{v}_t\mid\boldsymbol{u}_{t+1}}^{(i)}\propto \boldsymbol{w}_{\boldsymbol{v}_t}^{(i)}p(\widetilde{\boldsymbol{u}}_{t+1}\mid\boldsymbol{v}_{t}^{(i)})\\
				&\text{choose }\widetilde{\boldsymbol{u}}_{t}\text{ with probability }\boldsymbol{w}_{\boldsymbol{v}_t\mid\boldsymbol{u}_{t+1}}^{(i)}
			\end{aligned}$
		}
	}
	
	\tcp{estimate of expectation}
	\For{t=1:T}{
		$\begin{aligned}\mathbb{E}[\boldsymbol{u}_t]&\approx \textstyle\sum_{i=1}^N \boldsymbol{w}_t^{(i)} \boldsymbol{u}_t^{(i)}\\
			\mathbb{E}[\log\boldsymbol{u}_t]&\approx \textstyle\sum_{i=1}^N \boldsymbol{w}_t^{(i)} \log\boldsymbol{u}_t^{(i)}\\
			\mathbb{E}[\log\boldsymbol{v}_t]&\approx \textstyle\sum_{i=1}^N \boldsymbol{w}_t^{(i)} \log\boldsymbol{v}_t^{(i)}\end{aligned}$
	}
	\caption{The MC version of Gam-Chain's state estimation.}
	\label{alg:gcm}
\end{algorithm}
The complexity of this Algo. \ref{alg:gcm} is $\mathcal{O}\left(L*T*(4*N*s_{\text{g}}+g+e+a)\right)$\footnote{'g' represents the calculation of the gamma function.}. Compared to Algo. \ref{alg:pro}, it differs at $2*N*s_{\text{g}}$ and $s_{\text{a}}$. In order to compare more fairly, when comparing performance at Sec. \ref{sec:perf}, for any algorithm that requires MC, we set the number of particles to the minimum value, i.e., $2$; when comparing accuracy at Sec. \ref{sec:acc}, we set the number of particles to a large enough value, i.e., $20$.

\subsubsection{LogN-Chain/MC}
Similar to Gam-Chain/MC, we can also define the MC version of Eq. \ref{eqn:logn} as shown in Algo. \ref{alg:lcm}.
\begin{algorithm}[!htbp]
	
	\SetAlgoLined
	\setcounter{AlgoLine}{2}
	
	\tcp{E Step}
	\tcp{forward sampling}
	
	\For{t=1:T}{\label{step:e2}
		\For{i=1:N}{
			$\begin{aligned}
				\boldsymbol{w}_{\boldsymbol{u}_t}^{(i)} &\propto \boldsymbol{w}_{\boldsymbol{u}_{t-1}}^{(i)} p(\boldsymbol{y}_t\mid\boldsymbol{u}_{t})
			\end{aligned}$\tcp{update weights}
		}
	}
	\tcp{backward smoothing}
	\For{t=1:T}{
		\For{i=1:N}{
			$\begin{aligned}			&\boldsymbol{w}_{\boldsymbol{u}_t\mid\boldsymbol{u}_{t+1}}^{(i)}\propto \boldsymbol{w}_{\boldsymbol{u}_t}^{(i)}p(\widetilde{\boldsymbol{u}}_{t+1}\mid\boldsymbol{u}_{t}^{(i)})\\
				&\text{choose }\widetilde{\boldsymbol{u}}_{t}\text{ with probability }\boldsymbol{w}_{\boldsymbol{u}_t\mid\boldsymbol{u}_{t+1}}^{(i)}
			\end{aligned}$
		}
	}
	
	\tcp{estimate of expectation}
	\For{t=1:T}{
		$\begin{aligned}
			\mathbb{E}[\log^2\boldsymbol{u}_t]&\approx \textstyle\sum_{i=1}^N \boldsymbol{w}_t^{(i)} \log^2\boldsymbol{u}_t^{(i)}\\
			\mathbb{E}[\log\boldsymbol{u}_t\log\boldsymbol{u}_{t-1}]&\approx \textstyle\sum_{i=1}^N\textstyle\sum_{j=1}^N\boldsymbol{w}_t^{(i)}\boldsymbol{w}_{t-1}^{(j)} \log\boldsymbol{u}_t^{(i)}\log\boldsymbol{u}_{t-1}^{(j)}
		\end{aligned}$
	}
	\BlankLine
	\tcp{M Step}
	$S^2\leftarrow 1/T \left(\textstyle\sum_{i=2}^T \mathbb{E}[\log^2\boldsymbol{u}_t]-2 \mathbb{E}[\log\boldsymbol{u}_t\log\boldsymbol{u}_{t-1}]+\mathbb{E}[\log^2\boldsymbol{u}_{t-1}]\right)$
	
	\caption{The MC version of LogN-Chain's EM process.}
	\label{alg:lcm}
\end{algorithm}
Its complexity is the same as Algo. \ref{alg:gcm}, the difference is that only '$\exp$' needs to be calculated in Line \ref{step:e2} instead of $\Gamma$. However, note that $\Gamma(A)$ is the same for every EM iteration and it can be extracted outside the loop \ref{step:mloop}-\ref{step:mloop_end} in Algo. \ref{alg:pro}, so there should be no substantial difference in performance between them.

\subsubsection{LogN-Chain/VI}
Although the posterior of Eq. \ref{eqn:logn} has no closed form, it is still possible to approximate it with LA \cite{Kleppe2012FittingGS}.
\begin{algorithm}[!htbp]
	
	\SetAlgoLined
	\setcounter{AlgoLine}{2}
	
	\tcp{E Step}
	\For{t=1:T}{
		$\begin{aligned}\boldsymbol{\mu}_{t}&\leftarrow \frac{S^2}{4}+\frac{\boldsymbol{\mu}_{t-1}}{2}+\frac{\boldsymbol{\mu}_{t+1}}{2}-W\left(\frac{1}{4} S^2 \boldsymbol{y}_t^2 e^{\frac{S^2}{4}+\frac{\boldsymbol{\mu}_{t-1}}{2}+\frac{\boldsymbol{\mu}_{t+1}}{2}}\right)\\
			\boldsymbol{\sigma}^2_{t}&\leftarrow\frac{2}{e^{\boldsymbol{\mu}_t} \boldsymbol{y}_t^2+4/S^2}\end{aligned}$\label{step:s4}
	}	
	\tcp{estimate expectations}
	\For{t=1:T}{
		$\begin{aligned}
			\mathbb{E}[\log^2\boldsymbol{u}_t]&=\boldsymbol{\mu}^2_t+\boldsymbol{\sigma}_t^2\\
			\mathbb{E}[\log\boldsymbol{u}_t]&=\boldsymbol{\mu}_t
		\end{aligned}$
	}
	
	\BlankLine
	\tcp{M Step}
	$S^2\leftarrow 1/T \left(\textstyle\sum_{i=2}^T \boldsymbol{\mu}_t^2+\boldsymbol{\sigma}_t^2-2 \boldsymbol{\mu}_t \boldsymbol{\mu}_{t-1}+\boldsymbol{\mu}_{t-1}^2+\boldsymbol{\sigma}_{t-1}^2\right)$\label{step:m4}
	
	\caption{The VI version of LogN-Chain's EM process.where $W$ at line \ref{step:s4} is Lambert W function.}
	\label{alg:lca}
\end{algorithm}
Specifically in the E-step of Algo. \ref{alg:lca}, we use $LogN(\boldsymbol{\mu}_{\boldsymbol{u}_{t}},\boldsymbol{\sigma}^2_{\boldsymbol{u}_{t}})$ distribution to approximate $p(\boldsymbol{u}_{t}|\boldsymbol{y}_{1:T})$. And, based on the assumption of mean-field, there is $\mathbb{E}[\log\boldsymbol{u}_t \log\boldsymbol{u}_{t-1} ]=\mathbb{E}[\log\boldsymbol{ u}_t]\mathbb{E}[\log\boldsymbol{u}_{t-1}]$. Therefore, the calculation of the M-step can be simplified to the expression at Line \ref{step:m4} in Algo. \ref{alg:lca}. Overall, Algo. \ref{alg:lca} has the same algorithmic complexity as Algo. \ref{alg:pro}, but is much slower than arithmetic operations due to the use of Lambert W functions (refer to Sec. \ref{sec:perf} for details) ).

For the convenience of discussion, when the above four algorithms are mentioned later, they are sorted and named as C1\footnote{i.e., the 1st combination.} (LogN-Chain/VI), C2 (LogN-Chain/MC), C3 (Gam-Chain/VI) ), C4 (Gam-Chain/MC). We need to pay special attention to \textbf{C3} as this is the primary method recommended in this paper.

\section{Experiments}

\subsection{Data}

For the datasets, we selected crypto\footnote{obtained from Binance exchange.}, Nasdaq\footnote{obtained with pandas\_datareader.}, and Forex\footnote{obtained from www.myfxbook.com, quoted in USD.} markets. Among them, the volatility of the cryptocurrencies is the most extreme, which is conducive to testing the state estimation capability of the model. Moreover, like Forex, it trades 24 hours daily, facilitating comparisons between resolutions. The Nasdaq market is not only huge in volume and high in stock diversity but also has no intra-day limits. This is consistent with Eq. \ref{eqn:frame} that the support of historical returns should be $(-\infty,\infty)$. In terms of time span, we directly selected thousands of periods before the day of the experiment; at the minute and hour level of which, the cryptocurrency includes extreme fluctuations such as the LUNA crash; at the day level, Nasdaq includes the COVID crash, which is representative.

In data preprocessing, we converted all raw closing prices into log. returns, whose properties are shown in Tab. \ref{tab:data}. When there are no transactions, i.e., volume=0, the corresponding data points are removed. These empty points are meaningless in business. If they are not eliminated, the returns will not satisfy one continuous distribution but a mixture of zero and non-zero values, which is inconsistent with the basic assumption in Eq. \ref{eqn:frame}.

\begin{table}[htbp]
	\caption{Summary of Datasets. The 'Len.' in the header is the average length of the sequence (with no-transaction blanks removed). Define $\boldsymbol{r}=\Delta \log( \boldsymbol{\boldsymbol{price}})$, $\boldsymbol{v}=\Delta \log(\boldsymbol{r}^2)$. $\sigma_r$ represents the standard deviation of $\boldsymbol{r}$, $\gamma_r$ represents the kurtosis of $\boldsymbol{r}$, $\sigma_{v}$ represents the standard deviation of $\boldsymbol{v}$ , $\gamma_{v}$ represents the kurtosis of $\boldsymbol{v}$. The dataset D4 contains Nasdaq 100 Index constituents. The dataset D5 is a subset of Nasdaq by sorting tradable Nasdaq tickers alphabetically and picking the top 300.}
	\centering
	\begin{tabular}{@{}lllllllllll@{}}
		\toprule
		ID & Market & Start T.     & End T.     & \#Ins. & Freq. & Len.    & $\sigma_r$\tiny{($\times 10^{-3}$)} & $\gamma_r$ & $\sigma_{v}$ & $\gamma_{v}$ \\ \midrule
		D1 & crypto & 22-05-01 & 22-05-31 & 345 & 1m & 28463 & 4.146   & 82.16      & 1.893        & 4.646        \\
		D2 & crypto & 21-01-01 & 22-05-31 & 381 & 1h & 9350  & 20.79    & 91.48      & 2.851        & 3.743        \\
		D3 &
		crypto &
		17-08-17 &
		22-05-31 &
		406 &
		1d &
		573.1 &
		107.6 &
		29.77 &
		3.069 &
		3.893 \\
		D4 &
		nasdaq &
		17-01-03 &
		22-05-31 &
		102 &
		1d &
		1283.7 &
		59.06 &
		116.9 &
		3.173 &
		4.058 \\
		D5 &
		nasdaq &
		17-01-03 &
		22-05-31 &
		300 &
		1d &
		734.5 &
		72.58 &
		87.55 &
		2.777 &
		5.011 \\
		D6 & forex  & 22-05-20 & 22-05-31 & 27  & 1m & 15840 & 0.294   & 23.41      & 3.123        & 2.937        \\
		D7 & forex  & 22-01-01 & 22-05-31 & 27  & 1h & 3600  & 1.687   & 14.31      & 3.248        & 4.331        \\
		D8 & forex  & 20-01-01 & 22-05-31 & 27  & 1d & 881   & 7.409   & 3.857      & 3.154        & 3.783        \\ \bottomrule
	\end{tabular}	
\end{table}\label{tab:data}
From Tab. \ref{tab:data}, Nasdaq and Forex are indeed less volatile than cryptocurrencies. This is understandable because the tokens traded in the crypto market are neither stocks, guaranteed by future dividends, nor are they legal tender, endorsed by national credit. As a loosely defined "proof of stake," its value is often quite uncertain. In addition, although volatilities are unobservable, we can roughly estimate them point-by-point, shown in the last columns of Tab. \ref{tab:data}. It can be seen that the so-called "kurtosis of variance" should exist in crypto and stock markets (>3), and it seems even more significant in low-frequency data.

\subsection{Result of State Estimation}

\subsubsection{Distribution of Parameters}\label{sec:para}

This section will examine the distribution of parameters estimated by \textbf{C3} over different datasets. As the only parameter in the model, 'A' uniquely determines the following values: the kurtosis of returns $\gamma_{r}$, the variance of volatilities' increments $\sigma^2_{v}$, and the kurtosis of volatilities' increments $\gamma_{v}$. Regarding the relationship between A and the latter two, it can be seen from Sec. \ref{sec:dev_kurt} that both $\sigma^2_{v}$ and $\gamma_{v}$ are decreasing functions of A.

\begin{figure}[htbp]
	\centering	
	\begin{subfigure}[h]{0.45\textwidth}\label{fig:dist0}
		\includegraphics[width=\linewidth]{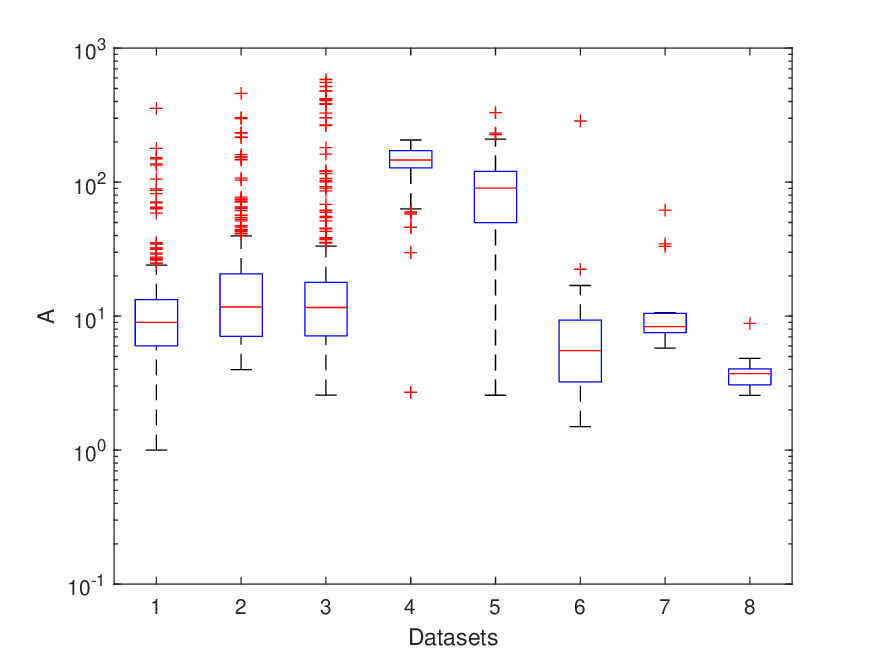}
		\caption{$\gamma_{r}$}
	\end{subfigure}
	\begin{subfigure}[h]{0.45\textwidth}\label{fig:dist1}
		\includegraphics[width=\linewidth]{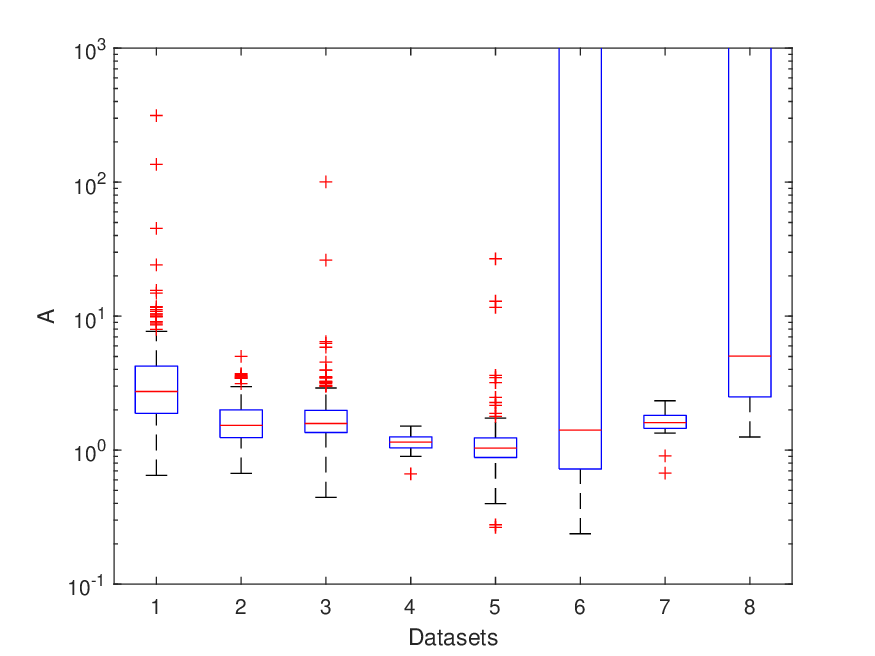}
		\caption{A}
	\end{subfigure}
	\caption{distribtions of $\gamma_{r}$ vs. distribtions of A}
	\label{fig:dist01}
\end{figure}

Next, we will study the relationship between $\gamma_{r}$ and A. Consider integrating out $\boldsymbol{u}_t$ in Eq. \ref{eqn:fix}, and let $\boldsymbol{v}_{t-1}\equiv B$ (i.e., remove the autocorrelation in volatilities), then get
\begin{equation}\label{eqn:tdist}
	p(\boldsymbol{y}_t\mid B)=\frac{2^A B^A  \left(2 B+\boldsymbol{y}_t^2\right)^{-A-\frac{1}{2}} \Gamma \left(A+\frac{1}{2}\right)}{\sqrt{\pi } \Gamma (A)}
\end{equation}
In fact, this is a non-standardized Student's t-distribution\footnote{Compound probability distribution, https://en.wikipedia.org/wiki/Compound\_probability\_distribution}. Its kurtosis is:
\begin{equation}
	\frac{3 \Gamma (A-2) \Gamma (A)}{\Gamma (A-1)^2}\text{ if }A>2
\end{equation}
This formula is also a decreasing function of A. To verify this, we separately trained the model of Eq. \ref{eqn:tdist} and compared its As with $\gamma_{r}$ in Tab. \ref{tab:data}, as shown in Fig. \ref{fig:dist01}.
In general, the larger the average $\gamma_{r}$ value of the data set, the smaller the corresponding A, and the inverse proportional relationship between them is generally validated.

\begin{figure}[htbp]
	\centering	
	\begin{subfigure}[h]{0.45\textwidth}\label{fig:dist2}
		\includegraphics[width=\linewidth]{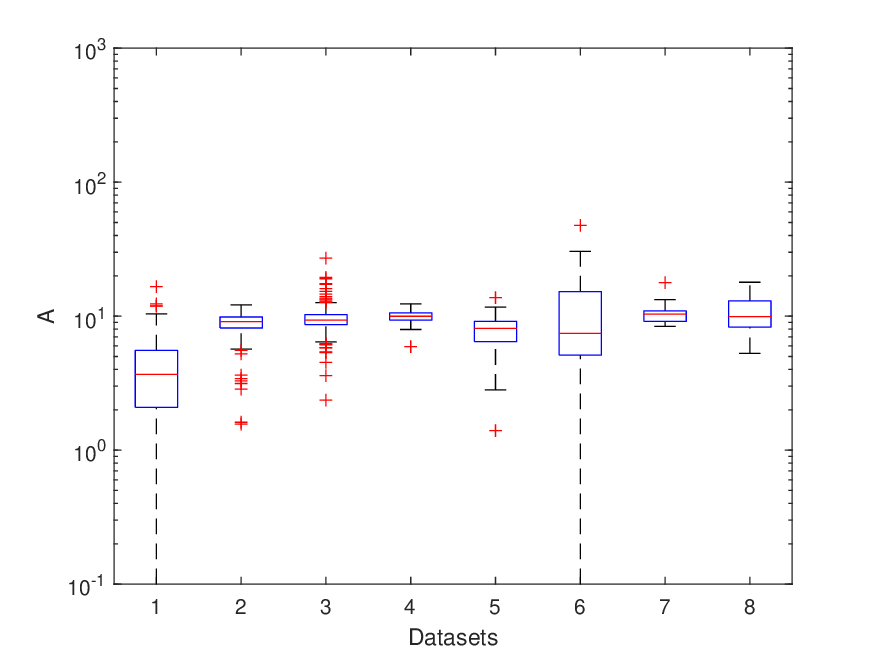}
		\caption{$\sigma^2_{v}$}
	\end{subfigure}
	\begin{subfigure}[h]{0.45\textwidth}\label{fig:dist3}
		\includegraphics[width=\linewidth]{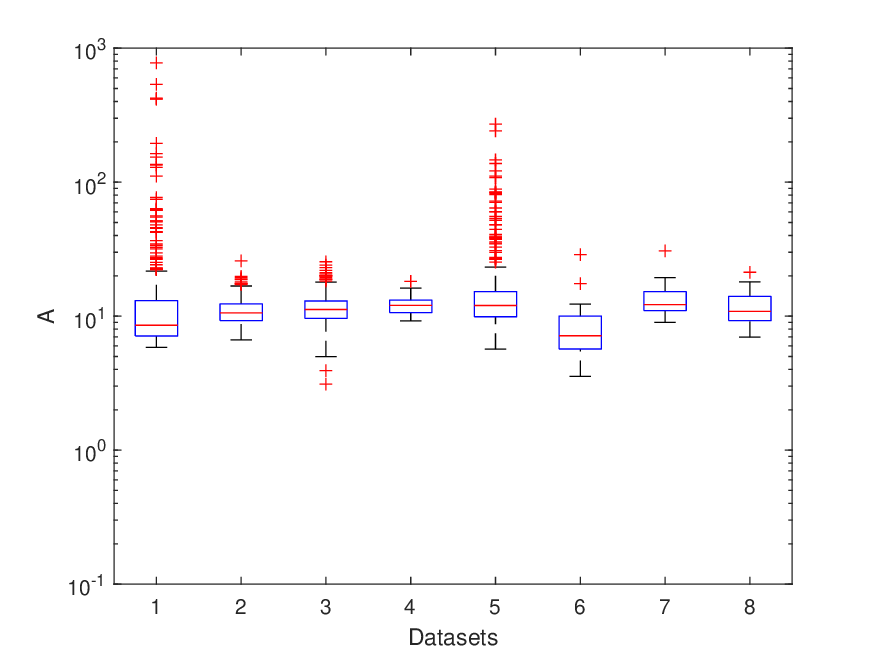}
		\caption{$\gamma_{v}$}
	\end{subfigure}
	\begin{subfigure}[h]{0.45\textwidth}\label{fig:dist4}
		\includegraphics[width=\linewidth]{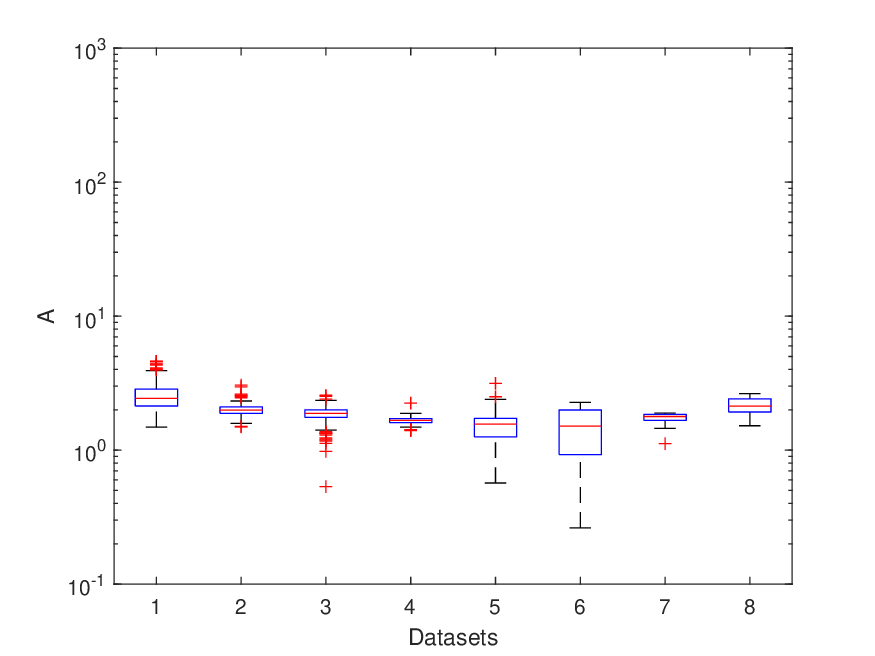}
		\caption{A}
	\end{subfigure}
	\caption{distribtions of $\sigma^2_{v}$, $\gamma_{v}$ vs. distribtions of A}
	\label{fig:dist234}
\end{figure}
Further, we run \textbf{C3} on all datasets and get the empirical distribution of A as shown in Fig. \ref{fig:dist234}(c). By comparing with Fig. \ref{fig:dist01}(b), we find that the autocorrelation between volatilities does not strongly impact the model estimation. Compared to Fig. \ref{fig:dist234}(c) with Fig. \ref{fig:dist234}(a), (b), we cannot find a very clear correlation yet. However, possibly due to the introduction of two layers of noise, the model of Eq. \ref{eqn:fix2} estimates $\boldsymbol{u}_t$ more smoothly, which makes it difficult to estimate A too large or too small. If it is too large, $\boldsymbol{u}_t$ will be nearly equal; if it is too small, it will cancel the volatility aggregation. These situations are both difficult to occur in practice; thus, the Eq. \ref{eqn:fix2} gives a more concentrated range of estimates than the Eq. \ref{eqn:tdist}.

\subsubsection{Residual Test}\label{sec:acc}

The state estimation results of C1-C4 are compared below. If $\boldsymbol{u}_t$ is observable, such as $\boldsymbol{u}_t^*$, then all residuals $\left\{\boldsymbol{e}_t\triangleq \boldsymbol{y} _t/\boldsymbol{u}_t^*\right\}$ will be exactly the standard normal distribution, which is so-called 'normalization'. However, since $\boldsymbol{u}_t$ is unobservable, we only have the posterior $p(\boldsymbol{u}_t\mid \boldsymbol{y}_{1:T})$, thus a workaround is, respectively sampling from $p(\boldsymbol{u}_t\mid \boldsymbol{y}_{1:T})$ to obtain $\boldsymbol{u}_t^s$, and then generating a residual set $\left\{\boldsymbol{e}_t^s\right\}$. The more accurate $p(\boldsymbol{u}_t\mid\boldsymbol{y}_{1:T})$ has estimated, the higher the probability $\left\{\boldsymbol{e}_t^s\right\}$ will pass the standard normal distribution test.Here we use the Kolmogorov-Smirnov test, which, as a nonparametric method, compares the difference between the empirical cumulative distribution and the $N(0,1)$'s cumulative distribution. This approach is intuitive and mimics the manual inspection we do in the Q-Q plot.

\begin{table}[htbp]
	\caption{Percentage of residuals for each instrument which can pass the KS-test.}
	\centering
	\begin{tabular}{@{}llllll@{}}
		\toprule
		& $\Delta \boldsymbol{y}_t$ & C1     & C2                               & \textbf{C3} & C4                               \\ \midrule
		D1 & 0                         & 0.1246 & \textbf{0.8260} & 0.6869                       & 0.7913                           \\
		D2 & 0                         & 0.1312 & 0.9632                           & 0.9160                       & \textbf{0.9685} \\
		D3 & 0.0073                    & 0.4778 & 0.9679                           & 0.9088                       & \textbf{0.9852} \\
		D4 & 0                         & 0.0196 & 0.8333                           & 0.7745                       & \textbf{0.8627} \\
		D5 & 0                         & 0.1652 & \textbf{0.9449} & 0.8601                       & 0.8986                           \\
		D6 & 0.2352                    & 0.8823 & \textbf{1}      & 0.9411                       & \textbf{1}      \\
		D7 & 0                         & 0      & \textbf{0.9411} & 0.8823                       & 0.8823                           \\
		D8 & 0                         & 0.7647 & 0.8823                           & 0.8235                       & \textbf{0.9411} \\ \bottomrule
	\end{tabular}
\end{table}\label{tab:normal}
We run the four algorithms on all datasets and check whether the residuals of each sequence are standard Gaussian, shown in Tab. \ref{tab:normal}. It can be seen that the MC-based algorithm still gets the best results, and whether gamma or lognormal is used, the results are very similar. However, as far as VIs are concerned, there are distinct differences. Based on the VI of Gam-chain, we get results that are 5\%-10\% worse than MC, yet basically acceptable in practice. The simple use of LA, due to the thin tail of the Gaussian distribution, is always worse than Gam-Chain.

\begin{figure}[htbp]
	\centering
	\begin{subfigure}[h]{0.48\textwidth}\label{fig:qq1}
		\includegraphics[width=\linewidth]{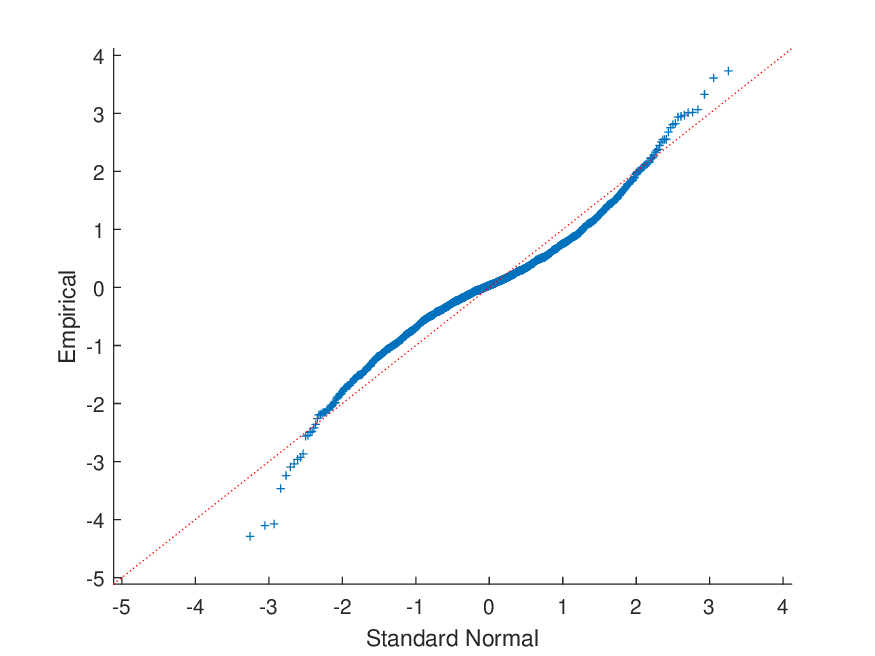}
		\caption{$C1$}
	\end{subfigure}
	\begin{subfigure}[h]{0.48\textwidth}\label{fig:qq2}
		\includegraphics[width=\linewidth]{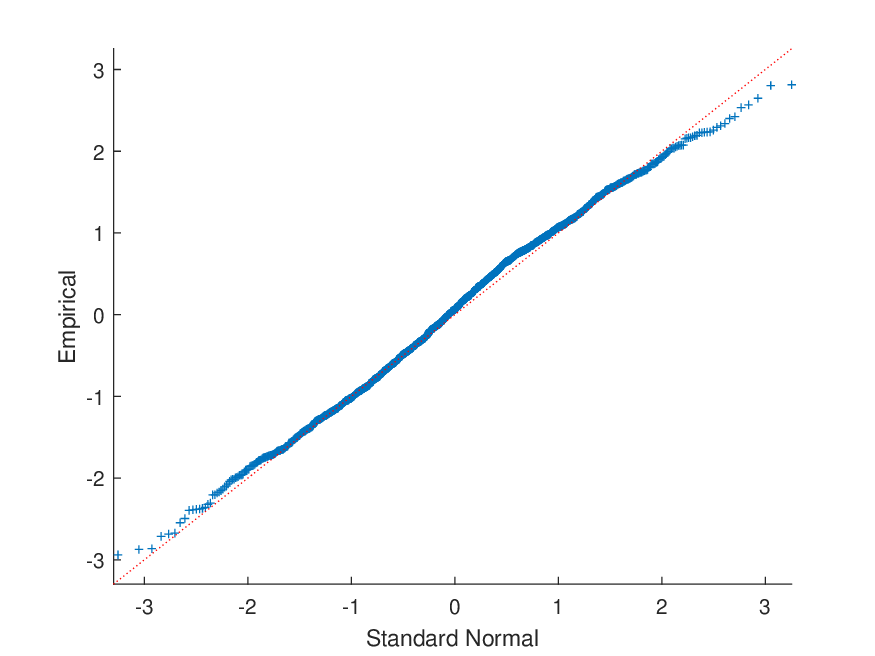}
		\caption{$C3$}
	\end{subfigure}
	\caption{QQ-plots of residuals generated by C1 and \textbf{C3}. The sequence is Tesla's day-level data, selected from dataset D4.}
\end{figure}\label{fig:qq}
We can also make an intuitive comparison for the effect of C1 and \textbf{C3} under the VI method. Fig. \ref{fig:qq} gives a Q-Q plot of the 'normalized' residuals for a specific sequence under both methods. By comparing the quantiles with the standard normal distribution, it can be seen that C1 underestimates the fluctuation in the tail and overestimates the fluctuation around the mean; in contrast, the difference between \textbf{C3} and the standard normal distribution is much smaller.

\subsection{Performance Comparison}\label{sec:perf}

Next, we will compare the performance of C1-C4 under different sequence lengths\footnote{The environment configuration is as follows. CPU: Intel64 Family 6 Model 142 Stepping 9 GenuineIntel ~2803 Mhz; Memory: 16,223 MB; OS: Win10; Compiler: MSVC 14.16; Additional Dependencies: boost 1.79.0}. If we change the specific instrument in testing, the impact on performance is minimal, so we must choose a long enough sequence to test.

\subsubsection{Running Time of Used Functions}

As we described in Sec. \ref{sec:var}, the key to performance is the functions used in the E-step. We tested the running time of the functions used in four algorithms, as shown in Tab. \ref{tab:functime}\footnote{This is implemented in C++. It does not require a virtual machine like Java or Python. It can directly use pointers (addresses) to read array elements, which is very efficient and removes the overhead of address translation. This allows us to have a more precise assessment of algorithm performance.}.
\begin{table}[htbp]
	\caption{The elapsed time of the function being used. The test method is to generate $10^9$ random numbers from $LogN(0,1)$ and take their average running time.}
	\centering
	\begin{tabular}{@{}lllllllll@{}}
		\toprule
		$10^{-9}$s & $+/-$    & $\times / \div$    & $e^x$    & $\log(x)$    & $x^y$  & $\Gamma(x)$   & $\psi(x)$ & $W(x)$ \\ \midrule
		Time   & 3.926 & 7.875 & 26.65 & 29.04 & 67.75 & 109.9 & 561.1 & 426.3   \\ \bottomrule
	\end{tabular}
\end{table}\label{tab:functime}
From the table, it can be inferred that, because of the different functions required, according to the discussion in the \ref{sec:var} section, the operation time order of the E-step should be: \textbf{C3}<C2<C4<C1. In M-step, the order of operation time is C1<C2<\textbf{C3}<C4. However, the extra time of \textbf{C3} in the M step caused by the $\psi(x)$ function is limited, for the number of calculations of $\psi(x)$ is fixed to 1 in each iteration. Therefore, it can be expected that when the sequence length $T$ increases, the consumption of \textbf{C3} on the $\psi(x)$ will be covered by the advantages obtained by the E-step, and it will run faster than any other method.

\subsubsection{Actual Measurement}

The time consumption of each step is tested below under different sequence lengths. For fairness, we have fixed the number of iterations to 1000. (It has been observed that there is no significant difference in the number of iterations for C1-C4. Moreover, the number of iterations for a longer sequence may not necessarily be more. Therefore, it is feasible to set a fixed value.) The test results are shown in the Fig. \ref{fig:perf}.
\begin{figure}[htbp]
	\centering
	\includegraphics[width=0.5\linewidth]{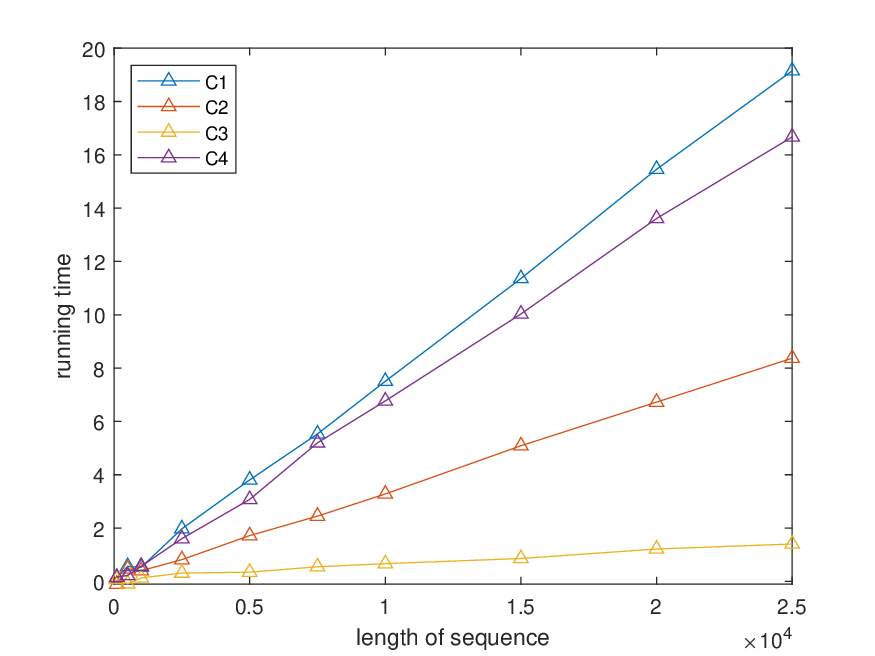}
	\caption{Comparison of time consumed. Here BTC/USDT is selected from D1 as the test sequence. It is truncated when a different sequence length is required. For the MC method, we only set the number of particles to 2.}
\end{figure}\label{fig:perf}
The results show that the time of the E-step almost dominates the increase of the running time as the sequence length grows. To be precise, \textbf{C3} only has a growth rate of about 20\% of the commonly used C2. This verifies the effectiveness of the scheme in this paper. In fact, the number of particles in C2 will not be set to 2 in practice, but to a larger number, such as 10. In this way, the speedup of \textbf{C3} will be multiplied, such as 20\%/5=4\%. Moreover, its calculation process is deterministic, unlike the MC method, which requires additional iterations to determine whether or not it has converged.
	
\section{Conclusion}
	
This paper presented an alternative scheme for estimating stochastic volatility based on the variational method. It can quickly estimate volatility for a large number of series, and the calculation process is entirely deterministic, so the convergence is also easy to judge. This is of particular practical value for high-frequency trading. In fact, the VI and MC methods are independent, and it is possible to perform fast initialization for any MC method using our method. Compared with the LA that belongs to the VI-class methods, the approximate posterior tail obtained by LA is heavier and does not need to calculate the Lambert W function. Therefore, it can improve both accuracy and performance.
	
In the future, we will consider introducing a more complex gamma network that can express the autocorrelation effect between volatility; or using multiple layers of latent gamma variables to achieve a broader range of kurtosis representation. In addition, this scheme has further room for optimization, such as using Taylor expansion to quickly calculate the digamma function, which will further improve the performance.

\appendix

\section{Details of Derivation}

\subsection{Derivation of the Distribution of Volatility Increments}\label{sec:dev_inc}

For the density function of the random variable function $Y=h(X)$, the general formula is
\begin{equation}\label{eqn:fun}
	g(y)=f\left(h^{-1}(y)\right)|\left(h^{-1}(y) \right )'|
\end{equation}
where $f(\cdot)$ is the density function of the random variable $X$.

For \ref{eqn:naive}, its $f(\cdot)$ is a gamma distribution. Since $\boldsymbol{w}_t\triangleq \Delta\log(\boldsymbol{u}_t)$, then $ \boldsymbol{u}_t=\boldsymbol{u}_{t-1}e^{\boldsymbol{w}_t}$. Substitute it into \ref{eqn:fun} to get
\begin{equation}\label{eqn:fun1}
	p(\boldsymbol{w}_t) = \frac{\boldsymbol{u}_{t-1}^A e^{\boldsymbol{u}_{t-1}^2 \left(-e^{\boldsymbol{w}_t}\right)} \left(\boldsymbol{u}_{t-1}e^{\boldsymbol{w}_t}\right)^{A-1}}{\Gamma (A)}  |\boldsymbol{u}_{t-1} e^{\boldsymbol{w}_t}|,
\end{equation}
 which is \ref{eqn:naive2} after simplification.

For \ref{eqn:fix}, first integrate out $\boldsymbol{v}_{t-1}$, the steps are as follows
\begin{equation}\label{eqn:fun2}
	\begin{aligned}
		p(\boldsymbol{u}_{t}|\boldsymbol{u}_{t-1})&=\int_0^{\infty} p(\mathbf{u}_{t}|\mathbf{v}_{t-1}) p(\mathbf{v}_{t-1}|\boldsymbol{u}_{t-1}) \text{d}\mathbf{v}_{t-1}\\
		&=\int_0^{\infty} \frac{\boldsymbol{u}_{t-1}^A \boldsymbol{u}_{t}^{A-1} \mathbf{v}_{t-1}^{2 A-1} e^{-\boldsymbol{u}_{t-1} \mathbf{v}_{t-1}-\boldsymbol{u}_{t} \mathbf{v}_{t-1}}}{\Gamma (A)^2} \text{d}\mathbf{v}_{t-1}\\
		&=\frac{\boldsymbol{u}_{t-1}^A \boldsymbol{u}_{t}^{A-1}(\boldsymbol{u}_{t-1}+\boldsymbol{u}_{t})^{-2 A}}{\Gamma (A)^2}\\
		&\cdot\int_0^{\infty} e^{-(\boldsymbol{u}_{t-1}+\boldsymbol{u}_{t})\boldsymbol{v}_{t-1}}\left((\boldsymbol{u}_{t-1}+\boldsymbol{u}_{t})\boldsymbol{v}_{t-1} \right)^{2A-1}(\boldsymbol{u}_{t-1}+\boldsymbol{u}_{t})\text{d}\mathbf{v}_{t-1}\\
		&=\frac{\boldsymbol{u}_{t-1}^A \boldsymbol{u}_{t}^{A-1}(\boldsymbol{u}_{t-1}+\boldsymbol{u}_{t})^{-2 A}}{\Gamma (A)^2}\Gamma (2A)\\
	\end{aligned}
\end{equation}
Similar to \ref{eqn:fun1}, we substitute $\boldsymbol{u}_t=\boldsymbol{u}_{t-1}e^{\boldsymbol{w}_t}$ and \ref{eqn:fun2} into \ref{eqn:fun}, and get
\begin{equation}\label{eqn:fun3}
	p(\boldsymbol{w}_t) = \frac{e^{(A-1) \boldsymbol{w}_{t}} \left(e^{\boldsymbol{w}_{t}}+1\right)^{-2 A} \Gamma (2 A)}{\cancel{\boldsymbol{u}_{t-1}} \Gamma (A)^2} |\cancel{\boldsymbol{u}_{t-1}} e^{\boldsymbol{w}_t}|,
\end{equation}
 which is \ref{eqn:fix2} after simplification. It can be seen from the above formula that $\boldsymbol{u}_{t-1}$ has been eliminated.

\subsection{Derivation of Moment Generating Function}\label{sec:dev_mgf}

In the model \ref{eqn:fix}, the moment generating function corresponding to $\boldsymbol{w}_t$ is
\begin{equation}
	\begin{aligned}
		\phi(\lambda)&=\int_0^{\infty}\int_0^{\infty} e^{\lambda  \log \frac{\boldsymbol{u}_t}{\boldsymbol{u}_{t-1}}} p(\boldsymbol{u}_t|\boldsymbol{v}_{t-1})p(\boldsymbol{v}_{t-1}|\boldsymbol{u}_{t-1}) \text{d}\boldsymbol{v}_{t-1} \text{d}\boldsymbol{u}_t\\
		&=\int_0^{\infty}\int_0^{\infty} \frac{\boldsymbol{v}_{t-1}^{2 A-1} \boldsymbol{u}_{t-1}^{A-\lambda} \boldsymbol{u}_t^{A+\lambda-1} e^{-\boldsymbol{v}_{t-1}(\boldsymbol{u}_{t-1}+\boldsymbol{u}_t)}}{\Gamma (A)^2} \text{d}\boldsymbol{v}_{t-1} \text{d}\boldsymbol{u}_t\\
		&=\frac{1}{\Gamma (A)^2}\int_0^{\infty}\left(\int_0^{\infty} (\boldsymbol{u}_t\boldsymbol{v}_{t-1})^{A+\lambda-1} e^{-\boldsymbol{u}_t\boldsymbol{v}_{t-1}}\boldsymbol{v}_{t-1}\text{d}\boldsymbol{u}_t\right)\\
		&\cdot(\boldsymbol{v}_{t-1}\boldsymbol{u}_{t-1})^{A-\lambda-1}e^{-\boldsymbol{v}_{t-1}\boldsymbol{u}_{t-1}}\boldsymbol{u}_{t-1}\text{d}\boldsymbol{v}_{t-1}\\
		&=\frac{1}{\Gamma (A)^2}\int_0^{\infty}\Gamma(A+\lambda)\cdot(\boldsymbol{v}_{t-1}\boldsymbol{u}_{t-1})^{A-\lambda-1}e^{-\boldsymbol{v}_{t-1}\boldsymbol{u}_{t-1}}\boldsymbol{u}_{t-1}\text{d}\boldsymbol{v}_{t-1}\\
		&=\frac{\Gamma(A+\lambda)\Gamma(A-\lambda)}{\Gamma (A)^2}\\
	\end{aligned}
\end{equation}
This is \ref{eqn:mgf}.

\subsection{Derivation of Kurtosis}\label{sec:dev_kurt}

After taking the derivative of \ref{eqn:mgf}, the 1st to 4th moments of $\boldsymbol{w}_t$ are obtained as:
\begin{equation}
	\begin{aligned}
		\mathbb{E}[\boldsymbol{w}_t]&=\phi'(0)=0,\\
		\mathbb{E}[\boldsymbol{w}_t^2]&=\phi''(0)=2 \psi^{(1)}(A),\\
		\mathbb{E}[\boldsymbol{w}_t^3]&=\phi'''(0)=0,\\
		\mathbb{E}[\boldsymbol{w}_t^4]&=\phi'''(0)=2 \left(6 \psi^{(1)}(A)^2 + \psi^{(3)}(A)\right)
	\end{aligned}
\end{equation}
Therefore, its variance and kurtosis are
\begin{equation}
	\begin{aligned}
		Var(\boldsymbol{w}_t)&=\mathbb{E}[\boldsymbol{w}_t^2] - \mathbb{E}[\boldsymbol{w}_t]^2\\
		&=2 \psi^{(1)}(A)\\
		K(\boldsymbol{w}_t)&=\frac{\mathbb{E}[\boldsymbol{w}_t]^4 - 4 \mathbb{E}[\boldsymbol{w}_t]^3 \mathbb{E}[\boldsymbol{w}_t] + 6 \mathbb{E}[\boldsymbol{w}_t]^2 \mathbb{E}[\boldsymbol{w}_t^2] - 4 \mathbb{E}[\boldsymbol{w}_t] \mathbb{E}[\boldsymbol{w}_t^3] + \mathbb{E}[\boldsymbol{w}_t^4]}{Var(\boldsymbol{w}_t)^2}\\
		&=3+\frac{\psi ^{(3)}(A)}{2 \psi ^{(1)}(A)^2}
	\end{aligned}
\end{equation}

\subsection{Proof of Kurtosis Bound}\label{sec:proof_kurt}

By the formula 6.4.10 on page 260 in \cite{abramovitz1968mathematical}, we know
\begin{equation}
	\psi^{(n)}(A)=(-1)^{n+1}n!\sum_{k=0}^{\infty}\frac{1}{(A+k)^{n+1}}.
\end{equation}
Then
$$
K(w_t)=3+\frac{\psi^{(3)}(A)}{2[\psi^{(1)}(A)]^2}=3+3\frac{\sum_{k=0}^{\infty}\frac{1}{(A+k)^{4}}}{[\sum_{k=0}^{\infty}\frac{1}{(A+k)^{2}}]^2}
$$
Since $[\sum_{k=0}^{\infty}\frac{1}{(A+k)^{2}}]^2=\sum_{k=0}^{\infty}\frac{1}{(A+k)^{4}}+\sum_{k\neq l}\frac{1}{(A+k)^{2}(A+l)^{2}}\geq \sum_{k=0}^{\infty}\frac{1}{(A+k)^{4}}$, then it is easy to see that
$
\frac{\sum_{k=0}^{\infty}\frac{1}{(A+k)^{4}}}{[\sum_{k=0}^{\infty}\frac{1}{(A+k)^{2}}]^2}\leq 1.
$
So $3\leq K(w_t)\leq 6$. Moreover, $\frac{\sum_{k=0}^{\infty}\frac{1}{(A+k)^{4}}}{[\sum_{k=0}^{\infty}\frac{1}{(A+k)^{2}}]^2}$ is a continuous function of $A$ on $(0,+\infty)$. When $A\to0$, we see $\sum_{k=0}^{\infty}\frac{1}{(A+k)^{4}}\sim \frac{1}{A^4}$ and $\sum_{k=0}^{\infty}\frac{1}{(A+k)^{2}}\sim \frac{1}{A^2}$. Therefore,
$$
\lim_{A\to0}\frac{\sum_{k=0}^{\infty}\frac{1}{(A+k)^{4}}}{[\sum_{k=0}^{\infty}\frac{1}{(A+k)^{2}}]^2}=1.
$$
When $A\to\infty$, we should approximate series $\sum_{k=0}^{\infty}\frac{1}{(A+k)^{4}}$ and $\sum_{k=0}^{\infty}\frac{1}{(A+k)^{2}}$ by improper integrals. Because $\frac{1}{(k+1+A)^2}\leq \int_{k}^{k+1}\frac{dx}{(x+A)^2}\leq \frac{1}{(k+A)^2}$ and  $\frac{1}{(k+1+A)^4}\leq \int_{k}^{k+1}\frac{dx}{(x+A)^4}\leq \frac{1}{(k+A)^4}$, then we know
\begin{align}
	\int_0^{\infty}\frac{dx}{(A+x)^4}\leq \sum_{k=0}^{\infty}\frac{1}{(A+k)^{4}}\leq \frac{1}{A^d}+\int_0^{\infty}\frac{dx}{(A+x)^4} \label{eq1}\\
	\int_0^{\infty}\frac{dx}{(A+x)^2}\leq \sum_{k=0}^{\infty}\frac{1}{(A+k)^{2}}\leq \frac{1}{A^d}+\int_0^{\infty}\frac{dx}{(A+x)^2}\label{eq2}
\end{align}
Then we know $\sum_{k=0}^{\infty}\frac{1}{(A+k)^{4}}\sim \int_0^{\infty}\frac{dx}{(A+x)^4}=\frac{1}{3A^3}$ and $\sum_{k=0}^{\infty}\frac{1}{(A+k)^{2}}\sim \int_0^{\infty}\frac{dx}{(A+x)^2}=\frac{1}{A}$ when $A\to\infty$. Thus,
$$
\lim_{A\to\infty}\frac{\sum_{k=0}^{\infty}\frac{1}{(A+k)^{4}}}{[\sum_{k=0}^{\infty}\frac{1}{(A+k)^{2}}]^2}=\lim_{A\to\infty}\frac{\frac{1}{3A^3}}{\frac{1}{A^2}}=0.
$$
Now we know $K(w_t)=3+\frac{\psi^{(3)}(A)}{2[\psi^{(1)}(A)]^2}$ is a continuous function of $A$. Both $K(w_t)=3$ and $K(w_t)=6$ are their horizontal asymptotic lines. Then $K(w_t)$ can assume all values in $(3,6)$.

\bibliographystyle{plain}
\bibliography{paper}  
	
\end{document}